\documentclass[12pt,preprint]{aastex}

\newcommand{\msun}{$M_{\odot}$}

\def\etal{{et al.\ }}

\def\zGP{$z_{\rm GP}$}

\def\Lya{{\rm Ly}$\alpha$}

\begin{document}

\title{Constraints on First-Light Ionizing Sources from 
Optical Depth of the Cosmic Microwave Background } 

\author{J. Michael Shull \& Aparna Venkatesan\altaffilmark{1} }

\affil{University of Colorado, Department of Astrophysical \&
     Planetary Sciences, \\ CASA, 389-UCB, Boulder, CO 80309 }

\altaffiltext{1}{Now at Department of Physics, 2130 Fulton St., 
   University of San Francisco, San Francisco, CA 94117}

\email{michael.shull@colorado.edu, avenkatesan@usfca.edu}

\begin{abstract}

We examine the constraints on high-redshift star formation, 
ultraviolet and X-ray pre-ionization, and the epoch of reionization 
at redshift $z_r$, inferred from the recent WMAP-5 measurement, 
$\tau_e = 0.084 \pm 0.016$, of the electron-scattering optical 
depth of the cosmic microwave background (CMB).  Half of this 
scattering can be accounted for by the optical depth, $\tau_e =$ 
0.04--0.05, of a fully ionized intergalactic medium (IGM) at 
$z \leq z_{\rm GP} \approx$ 6--7, consistent with Gunn-Peterson 
absorption in neutral hydrogen.  
The required additional optical depth, $\Delta \tau_e = 0.03 \pm 0.02$
at $z >$ \zGP, constrains the ionizing contributions of 
``first light" sources.  WMAP-5 also measured a significant increase in 
small-scale power, which lowers the required efficiency
of star formation and ionization from mini-halos.
Early massive stars (UV radiation) and black holes (X-rays) can 
produce a partially ionized IGM, adding to the residual electrons 
left from incomplete recombination.  Inaccuracies in computing the 
ionization history, $x_e(z)$, and degeneracies in cosmological 
parameters ($\Omega_m$, $\Omega_b$, $\sigma_8$, $n_s$) add systematic 
uncertainty to the measurement and modeling of $\tau_e$.  From the 
additional optical depth from sources at $z >$ \zGP, we limit the 
star-formation efficiency, the rate of ionizing photon 
production for Pop~III and Pop~II stars, and the photon escape fraction, 
using standard histories of baryon collapse, minihalo star formation, 
and black-hole X-ray preionization.  

\end{abstract}


\clearpage
\newpage

\section*{1. INTRODUCTION }

In recent years, an enormous amount of exciting cosmological data 
have appeared, accompanied by theoretical inferences about early 
galaxy formation and the first massive stars.  Many of these 
inferences were reactions to first-year (WMAP-1) results (Kogut et al.\
2003; Spergel \etal 2003) from the {\it Wilkinson Microwave Anisotropy 
Probe} (WMAP).  WMAP-1 inferred a high optical depth to the cosmic 
microwave background (CMB) and suggested early reionization of the 
intergalactic medium (IGM).  Other conclusions came from simplified 
models for the stellar initial mass function (IMF), 
atomic/molecular physics, radiative processes, and prescriptions for 
star formation rates and escape of photoionizing radiation from
protogalaxies. 

The CMB optical depth and other cosmological parameters have been
refined significantly in the recent WMAP-5 data (Hinshaw \etal 2008).
In this paper, we use these new measurements to constrain the efficiency
of first-light ionizing sources.  We focus on the reionization epoch, defined 
as the redshift $z_r$ when the IGM becomes nearly fully ionized over most of 
its volume (Gnedin 2000, 2004).  Our knowledge about reionization comes 
primarily from three types of observations:  hydrogen \Lya\ absorption 
in the IGM, high-$z$ \Lya-emitting galaxies, and CMB optical depth.
Optical spectroscopic studies of the ``Gunn-Peterson" (\Lya) 
absorption toward high-redshift quasars and galaxies imply that H~I 
reionization occurred not far beyond $z_{\rm GP} \sim 6$ 
(Becker \etal 2001; Fan \etal 2002, 2006).  Ultraviolet spectra suggest
that He~II reionization occurred at $z \sim 3$ (Kriss \etal 2001;
Shull \etal 2004; Zheng \etal 2004).  The detection of high-redshift 
(\Lya-emitting) galaxies
(Hu \& Cowie 2006) suggests a somewhat higher redshift, $z_r \geq 6.5$. 
Data from WMAP, after three years (Spergel \etal 2007) and five years 
(Hinshaw \etal 2008), suggest that reionization might occur at $z \approx 10$, 
with sizeable uncertainties in measuring and modeling the CMB optical depth.  

There appears to be a discrepancy between the two epochs, 
\zGP\ $\approx$ 6--7 and $z_r \approx 10$.  
However, the WMAP and \Lya\ absorption results are not necessarily
inconsistent, since they probe small amounts of ionized and neutral gas, 
respectively.  Both the H~I absorbers and ionized filaments 
in the ``cosmic web" (Cen \& Ostriker 1999) are highly structured 
at redshifts $z < 10$ and affect the optical depths in \Lya.  
In order to effectively absorb all the \Lya\ radiation 
at $z \approx 6$ requires a volume-averaged neutral fraction of just 
$x_{\rm HI} \approx 4 \times 10^{-4}$ (Fan \etal 2006).
Simulations of the reionization process (Gnedin 2004; Gnedin \& Fan 2006) 
show that the transition from neutral to ionized is extended in time between 
$z = 5-10$.  The first stage (pre-overlap) involves the development and 
expansion of the first isolated ionizing sources. The second stage marks 
the overlap of the ionization fronts and the disappearance of the last 
vestiges of low-density neutral gas.  Finally, in the post-overlap stage, 
the remaining high-density gas is photoionized.  The final
drop in neutral fraction and increase in photon mean free path occur quite 
rapidly, over $\Delta z \approx 0.3$.

Initial interpretations of the high WMAP-1 optical depth 
($\tau_e = 0.17 \pm 0.03$) were based on models of ``sudden reionization", 
neglecting scattering from a partially ionized IGM at $z > z_r$. The 
three-year (WMAP-3) and five-year (WMAP-5) data resulted in considerable 
downward revisions, with recent measurements giving
$\tau_e = 0.084 \pm 0.016$ (Hinshaw \etal 2008; Komatsu \etal 2008). 
These analyses assumed a single step to complete ionization ($x_e = 1$ at 
$z \leq z_r$), while Dunkley \etal (2008) also explored a two-step 
ionization history with a broader distribution of $\tau_e$. 
The WMAP-5 data alone gave a $5 \sigma$ detection, 
$\tau_e = 0.087 \pm 0.017$ (Dunkley \etal 2008), while WMAP-5 combined
with distance measurements from Type~Ia supernovae (SNe) and baryon 
acoustic oscillations (BAO) gave $\tau_e = 0.084 \pm 0.016$ 
(Hinshaw \etal 2008; Komatsu \etal 2008).    
The single-step analyses gave a reionization epoch 
$z_r = 10.8 \pm 1.4$ at 68\% confidence level (C.L.). 
The $\chi^2$ curves for two-step ionization provide little 
constraint on the redshift of reionization or on an IGM with low partial 
ionization fractions, $x_e \ll 0.1$, as we discuss in \S~3. 
Figure 3 of Spergel \etal (2007) and Figure 8 of Dunkley \etal (2008) 
show that the parameters $x_e$ and $z_r$ are somewhat degenerate, each with 
a long tail in the likelihood curves.  This emphasizes the importance of contributions 
to $\tau_e$ from ionizing UV photons at $z >$ \zGP\ from early massive stars 
and X-rays from accreting black holes (Venkatesan, Giroux, \& Shull 2001,
hereafter VGS01; Ricotti \& Ostriker 2004; Begelman, Volonteri, \& Rees 2006).

Models of the extended recombination epoch (Seager, Sasselov, \& Scott 2000)
predict a partially ionized medium at high redshifts, owing to residual
electrons left from incomplete recombination.  Residual electrons produce 
additional scattering, $\Delta \tau_e \approx 0.06$ between redshifts
$z \approx$ 10--700.  
These effects are computed in CMB radiation transfer codes such as 
CMBFAST and RECFAST, but only a portion of this scattering affects the large 
angular scales ($\ell \leq 10$) where WMAP detects
a polarization signal.  X-ray preionization can also produce CMB
optical depths $\tau_e \geq 0.01$ (VGS01; Ricotti, Ostriker, \& Gnedin 2005).  

As we discuss in \S~2.1, a fully ionized IGM from $z$ = 0 back to 
\zGP\ $= 6.1 \pm 0.15$, the reionization epoch inferred from Gunn-Peterson 
absorption (Gnedin \& Fan 2006), produces optical depth, 
$\tau_e \approx 0.041 \pm 0.003$.  This represents nearly half of 
the WMAP-5 measurement, $\tau_e = 0.084 \pm 0.016$.  
If the epoch of Gunn-Peterson reionization is \zGP\ = 7 
(Wyithe \etal 2008), the contribution is $\tau_e = 0.050 \pm 0.003$.   
Therefore, the high-redshift ionizing sources are limited to 
producing an additional optical depth, $\Delta \tau_e \leq 0.03 \pm 0.02$. 
In \S~3, we discuss the resulting constraints on the amount of star formation 
and X-ray activity at $z \ga 7$.  We also parameterize the efficiency of
various star formation environments, as determined by halo mass and  
the metallicity of stellar populations, to address several questions.  
Can reionization occur at relatively
low efficiency, corresponding to star formation in massive \Lya-cooled halos
(virial temperature $T_{\rm vir} \approx 10^4$ K), or must it take place 
in H$_2$-cooled minihalos ($T_{\rm vir} \approx 10^3$ K)?  Does one
require the high ionizing efficiencies of Population~III (metal-free) stars,
or can reionization be achieved by lower-efficiency Population~II stars?   
 
The IGM is thought to have a complex reionization history, with periods of 
extended reionization for H~I and He~II (Venkatesan, Tumlinson, \& Shull 2003; 
Cen 2003; Wyithe \& Loeb 2003; Hui \& Haiman 2003; Benson \etal 2006).  
Because $\tau_e$ measures the integrated column density of electrons, there 
are many possible scenarios consistent with the WMAP-5 data.  In addition, 
reionization models are quite sensitive to the amount of ``small-scale power"
for ionizing sources, which depend on $\sigma_8$, the normalized amplitude of
fluctuations.

\section*{2. OPTICAL DEPTH TO ELECTRON SCATTERING }

\subsection*{2.1. Analytic Calculation of Optical Depth} 

To elucidate the dependence of CMB optical depth on the epoch of reionization 
(redshift $z_r$), we integrate the electron scattering optical depth,
$\tau_e(z_r)$, for a homogeneous, fully ionized medium out to $z_r$.  
For instantaneous, complete ionization at redshift $z_r$, we calculate 
$\tau_e$ as the integral of $n_e \sigma_T d \ell$, the 
electron density times the Thomson cross section along proper length,   
\begin{equation}
   \tau_e(z_r) = \int _{0}^{z_r} n_e \sigma_T (1+z)^{-1} \; [c/H(z)] \; dz \; .   
\end{equation} 
We adopt a standard $\Lambda$CDM cosmology, in which 
$(d \ell /dz) = c(dt/dz) = (1+z)^{-1} [c/H(z)]$, where
$H(z) = H_0 [\Omega_m (1+z)^3 + \Omega_{\Lambda}]^{1/2}$ and
$\Omega_m + \Omega_{\Lambda} = 1$ (no curvature).  
The densities of hydrogen, helium, and electrons are written
$n_H = [(1-Y) \Omega_b \rho_{\rm cr}/m_H](1+z)^3$, $n_{\rm He} = y n_H$,
and $n_e = n_H (1+y)$, if helium is singly ionized.  
We assume a primordial helium mass fraction,
$Y = 0.2477\pm0.0029$ (Peimbert \etal 2007) and define 
$y = (Y/4)/(1-Y) \approx 0.0823$ as the He fraction by number.  
Helium contributes 8\% to $\tau_e$, assuming single ionization (He~II) 
at $z > 3$. An additional $\tau_e \approx 0.002$ comes from helium
reionization to He~III at $z \leq 3$ (Shull \etal 2004).  
The critical density is 
$\rho_{\rm cr} = (1.8785 \times 10^{-29}$~g~cm$^{-3}) h^2$
where $h = (H_0$/100~km~s$^{-1}$~Mpc$^{-1}$). 
The above integral can be done analytically:
\begin{eqnarray}
   \tau_e(z_r) &=& \left( \frac {c}{H_0} \right) \left( \frac {2 \Omega_b}
      {3 \Omega_m} \right)
     \left[ \frac {\rho_{\rm cr} (1-Y)(1+y) \sigma_T } {m_H} \right]
     \left[ \{ \Omega_m (1+z_r)^3 + \Omega_{\Lambda} \}^{1/2} - 1 \right] 
            \;  \nonumber \\  
    &=& (0.00435) h_{70} \left[ \{ 0.2794 h_{70}^{-2} (1+z_r)^3 
        + 0.7206 \} ^{1/2} - 1 \right] \; ,  
\end{eqnarray}
where we write $(1-Y)(1+y) = (1-3Y/4)$ and use the updated WMAP-5 parameters,
$\Omega_b h^2 = 0.02265 \pm 0.00059$ and $\Omega_m h^2 = 0.1369 \pm 0.0037$. 
To this formula we add the extra scattering ($\tau_e \approx 0.0020$) from 
He~III at $z \leq 3$.  With this He~III contribution, 
equation (2) yields $\tau_e =$ 0.040, 0.045, and 0.050 for 
$z_r =$ 6.0, 6.5, and 7.0, respectively.   

For large redshifts, $\Omega_m (1+z)^3 \gg \Omega_{\Lambda}$, and the 
integral simplifies to
\begin{equation}
  \tau_e(z_r) \approx \left( \frac {c}{H_0} \right) \left( \frac {2 \Omega_b}
   {3 \Omega_m^{1/2}} \right)  \left[ \frac {\rho_{\rm cr} (1-3Y/4) \sigma_T}
   {m_H} \right] (1+z_r)^{3/2} \approx (0.0522) \left[ \frac
    {(1+z_r)}{8} \right] ^{3/2}  \; .   
\end{equation}
Here, we have scaled to a reionization epoch $z_r \approx 7$. 
From the approximate expression in equation (3), 
we see that $\tau_e \propto (\rho_{\rm cr} \Omega_b \Omega_m^{-1/2} H_0^{-1}$). 
Thus, $\tau_e$ is nearly independent of the Hubble constant, since
$\rho_{\rm cr} \propto h^2$ while the combined parameters, $\Omega_b h^2$ 
and $\Omega_m h^2$, are inferred from D/H, CMB, and large-scale galaxy motions. 
The scaling with $h$ therefore cancels to lowest order.  A slight dependence 
remains from the small $\Omega_{\Lambda}$ term in equation (2).

If we invert the approximate equation (3), we can estimate the redshift 
of primary (Gunn-Peterson) reionization, 
$(1+z_{\rm GP}) \approx (7.77) [\tau_e(z_{\rm GP})/0.05]^{2/3}$,
scaled to the value, $\tau_e = 0.05$, expected for full
ionization back to $z_{\rm GP} \approx 7$.  As discussed in \S~2.2, 
this is approximately the WMAP-5 value of optical depth, 
$\tau_e = 0.084 \pm 0.016$, reduced by $\Delta \tau_e \approx 0.03$.  
This additional scattering, $\Delta \tau_e$, may arise from high-$z$ star 
formation, X-ray preionization, and residual electrons left after incomplete 
recombination.  The latter electrons are computed to have fractional ionization 
$x_e \approx (0.5-3.0) \times 10^{-3}$ between $z$ = 10--700 (Seager \etal 
2000).  Inaccuracies in computing their contribution therefore add
systematic uncertainty to the CMB-derived value of $\tau_e$.  
Partial ionization may also arise from the first stars 
(Venkatesan, Tumlinson, \& Shull 2003, hereafter VTS03) 
and from penetrating X-rays produced by early black holes (VGS01;
Ricotti \& Ostriker 2004, 2005).  For complete sudden reionization,
Komatsu \etal (2008) estimated $z_r \approx 10.8 \pm 1.4$ at 68\% C.L., 
by combining WMAP-5 data with other distance measures (SNe, BAO).   
The WMAP-5 data alone (Dunkley \etal 2008) imply $\tau_e = 0.087 \pm 0.017$,
with $z_r = 11.0 \pm 1.4$ (68\% C.L.). Their likelihood curves allow a range 
$7.5 \leq z_r \leq 13.0$ at 95\% C.L.  They also claim that WMAP-5 data 
exclude $z_r = 6$ at more than 99.9\% C.L.

The additional ionization sources at $z >$ \zGP\ will contribute electron 
scattering that may bring the WMAP and Gunn-Peterson results into agreement 
for the epoch of complete reionization. 
In our calculations, described in \S~3, we make several key assumptions.  
First, we assume a fully ionized IGM out to \zGP\ $\approx$ 6--7, accounting 
for both H$^+$ and ionized helium. 
Second, we investigate the effects of IGM partial ionization at $z >$ \zGP.
Finally, in computing the contribution of residual electrons at
high redshifts, we adopt the concordance parameters from the WMAP-5 
data set.  The CMB optical depth is formally a $5 \sigma$ result,
which may improve as WMAP refines its estimates of the matter density, 
$\Omega_m$, and the parameters, $\sigma_8$ and $n_s$, that 
govern small-scale power.  Both $\sigma_8$ and $n_s$ have well-known 
degeneracies with $\tau_e$ in CMB parameter extraction (Spergel \etal 2007;
Dunkley \etal 2008).  The derived parameters may change in future CMB 
data analyses, as the constraints on $\tau_e$
continue to evolve.  In addition, inaccuracies in the incomplete 
recombination epoch and residual ionization history, $x_e(z)$, add 
uncertainties to the CMB radiative transfer, the damping
of $\ell$-modes, and the polarization signal used to derive an
overall $\tau_e$.

\subsection*{2.2. Residual Electrons in the IGM}

We now discuss the contribution of residual electrons in the IGM following 
the recombination epoch at $z \approx 1000$.  Scattering from these
electrons is significant and is normally accounted for in CMB transport codes 
such as CMBFAST (Seljak \& Zaldarriaga 1996) through the post-recombination
IGM ionization history, $x_e(z)$.  However, a number of past papers are
vague on how the ionization history is treated and on precisely which
electrons contribute to total optical depth $\tau_e$. This has led to 
confusion in how much residual optical depth and damping of CMB power 
has been subtracted from the CMB signal.  It is important to be clear
on the definition of the integrated $\tau_e$ when using it to constrain
the amount of high-$z$ ionization from first-light sources.     
Modern calculations of how the IGM became neutral have been done 
by Seager \etal (2000), although their code (RECFAST) continues to be modified
to deal with subtle effects of the recombination epoch and the atomic
physics of hydrogen ($ns \rightarrow 1s$ and $nd \rightarrow 1s$) two-photon 
transitions (Chluba \& Sunyaev 2007). 

To illustrate the potential effects of high-$z$ residual electrons,
we have used numbers from Figure 2 of Seager \etal (2000), the 
top-panel model, which assumed a cosmology with 
$\Omega_{\rm tot} = 1$, $\Omega_b = 0.05$, $h = 0.5$, $Y = 0.24$, 
and $T_{\rm CMB} = 2.728$~K.  At low redshifts, $z \approx z_r$, 
just before reionization, they find a residual electron fraction 
$x_{e,0} \approx 10^{-3.3}$.  We fitted their curve for log~$x_e$ 
out to $z \approx 500$ to the formula:
\begin{equation}
  x_e (z) = x_{e,0} \; 10^{0.001(1+z)} \approx 
   (5 \times 10^{-4}) \; \exp [\alpha (1+z)] \; , 
\end{equation}
where $\alpha \approx 2.303 \times 10^{-3}$.  More recent recombination 
calculations (W. Y. Wong \& D. Scott, private communication) using WMAP 
parameters ($\Omega_b = 0.04$, $h = 0.73$, $\Omega_m = 0.24$, 
$\Omega_{\Lambda} = 0.76$) and $Y = 0.244$ find somewhat lower values, 
$x_{e,0} \approx 10^{-3.67}$ with $\alpha \approx 2.12 \times 10^{-3}$.  
The lower $x_{e,0}$ to the faster recombination rates arises
from their higher assumed baryon density, $\Omega_b h^2 = 0.0213$, compared 
to $\Omega_b h^2 = 0.0125$ in Seager \etal (2000).    
 
To compute the electron-scattering of the CMB from these ``frozen-out"
electrons, we use the same integrated optical depth formula (equation 1), 
in the high-$z$ limit, where 
$(d \ell /dz) = (1+z)^{-1} [c/H(z)] \approx (c/H_0) \Omega_m^{-1/2} (1+z)^{-5/2}$.  
We integrate over the residual-electron history, from $z_r \approx 7$ back 
to a final redshift $z_f \gg z_r$, to find
\begin{eqnarray}
 (\Delta \tau_e)_{\rm res} &=& \left( \frac {c} {H_0} \right) 
   \left[ \frac { \rho_{\rm cr} (1-Y) \sigma_T \Omega_b } 
        { \Omega_m^{1/2} m_H } \right]
      \int_{z_r}^{z_f} (1+z)^{1/2} \; x_e(z)  \; dz  \nonumber    \\   
     &=& (3.18 \times 10^{-3}) \; x_{e,o} \int_{(1+z_r)}^{(1+z_f)} 
         u^{1/2} \; \exp (\alpha u) \; du   \; .  
\end{eqnarray}
A rough estimate to the residual scattering comes from setting $\alpha = 0$
and adopting a constant ionized fraction $x_e(z)$, 
\begin{equation}
  (\Delta \tau_e)_{\rm res} \approx (2.12 \times 10^{-3})\; 
    \langle x_e \rangle \; [(1+z_f)^{3/2} - (1+z_r)^{3/2}] \; .
\end{equation}
This estimate gives $\tau_e = 0.039$ for $z_r = 6$, $z_f = 700$, and 
$\langle x_e \rangle \approx 10^{-3}$.   
More precise values of $\tau_e$ can be derived from the exact integral 
(eq.\ 5) by expanding the exponential as a sum and adopting the limit 
$z_f \gg z_r$,
\begin{equation} 
   (\Delta \tau_e)_{\rm res} = (0.0179) \left[ 
                 \frac {(1+z_f)} {501} \right]^{3/2} \; 
   \sum _{n=0}^{\infty} \frac { [\alpha (1+z_f)]^n }{ (n+3/2) \; n! } \; \; . 
\end{equation}

For $z > 500$, the approximate formula (eq.\ 4) underestimates $x_e$,
but one can integrate the appropriate curves (Seager \etal 2000)
using piecewise-continuous linear fits. Table 1 lists the fitting 
parameters, $x_{e,0}$ and $\alpha$, for various redshift ranges,
together with the extra contribution, $\Delta \tau_e$.   
These calculations give a total optical depth in residual electrons 
$\Delta \tau_e \approx 0.06$ back to $z = 700$.  These electrons
have maximum influence on angular scales with harmonic 
$\ell_{\rm max} \approx 2 z^{1/2} \approx$ 20--50 (Zaldarriaga 1997).  
At higher redshifts, $x_e$ rises to $10^{-2.1}$ at $z = 800$ and to
$10^{-1.1}$ at $z = 1000$, where the CMB source function will affect 
the ``free-streaming" assumption used in CMBFAST 
(Seljak \& Zaldarriaga 1996).    
 

\begin{deluxetable}{llll}
\tablecolumns{4}
\tablewidth{0pt}
\tablecaption{Ionization Fractions and Residual Optical Depth}
\tablehead{
        \colhead{Redshift Range}                   &
        \colhead{$x_{e,0}$ \tablenotemark{a} }            &
        \colhead{$\alpha$ \tablenotemark{a} }       &
        \colhead{$\Delta \tau_e$ \tablenotemark{b}}  }
\startdata
$7   < z < 500$ & $5.00 \times 10^{-4}$ & $2.303 \times 10^{-3}$ & 0.0256 \\
$500 < z < 600$ & $2.94 \times 10^{-4}$ & $3.224 \times 10^{-3}$ & 0.0135 \\
$600 < z < 700$ & $1.28 \times 10^{-4}$ & $4.606 \times 10^{-3}$ & 0.021  \\
\enddata
\tablenotetext{a}{Parameters for $x_e(z) = x_{e,0}\exp[-\alpha (1+z)]$ (see text)}
\tablenotetext{b}{Residual integrated optical depth computed over given
    redshift range (see eq.\ 5)}
\end{deluxetable}

\section*{3. IMPLICATIONS FOR REIONIZATION MODELS }

The WMAP-5 measurements  (Hinshaw \etal 2008; Komatsu \etal 2008)
of fluctuations in temperature ($T$) and polarization ($E$) have been 
interpreted to estimate {\it total} electron-scattering optical depth 
of $\tau_e \approx 0.084 \pm 0.016$.  The central value 
comes from computing the likelihood function for the six-parameter fit to 
WMAP-5 data (TT, TE, EE) marginalized with BAO and SNe distance meaasures. 
Because of the challenges in translating a single parameter ($\tau_e$)
into a reionization history, $x_e(z)$, it is important to recognize 
the sizable error bars on $\tau_e$. At 68\% C.L. 
(Figure 1 of Komatsu \etal 2008; Figure 6 of Dunkley \etal 2008), when 
marginalized against other parameters such as ``tilt" ($n_s$), the
optical depth ranges from $\tau_e = 0.06$ -- $0.11$.  The lower value is 
only slightly above the optical depth, $\tau_e = 0.05$, for a fully ionized
IGM back to redshift $z_{\rm GP} \approx 7$.  The higher value, 
$\tau_e = 0.11$, clearly requires additional ionizing sources at $z >$ \zGP.

In \S~2.1, we showed that $\sim50$\% of this $\tau_e$ can be accounted 
for by a fully ionized IGM at $z \leq$ \zGP.  Observations of \Lya\ 
(Gunn-Peterson) absorption toward 19 quasars between $5.7 < z < 6.4$ 
(Fan \etal 2006) are consistent with a reionization epoch 
$z_r \approx z_{\rm GP} = 6.1 \pm 0.15$ (Gnedin \& Fan 2006) . According to 
equation (2), this produces $\tau_e = 0.041 \pm 0.003$, where our error propagation 
includes relative uncertainties in $z_r$ (2.5\%), $\Omega_m h^2$ (8.4\%),
and $\Omega_b h^2$ (4.0\%).  Detections of \Lya\  
emitting galaxies at $z \approx 6.6$ (Hu \etal 2002; Kodaira \etal
2003; Hu \& Cowie 2006) suggest that the epoch of full reionization 
might be as high as $z_r \approx 7$.  This reionization epoch corresponds to 
$\tau_e \approx 0.05$ and is consistent with the small neutral fraction 
to which Gunn-Peterson test is sensitive, particularly when one accounts 
for density bias in the observed high-$z$ ionization zones around quasars 
(Wyithe, Bolton, \& Haehnelt 2008).   Residual post-recombination 
electrons produce a substantial optical depth from $z \approx 10$
back to $z \approx 700$, which uniformly damps all angular scales.
However, their effect on the TE and EE power is considerably
less on large angular scales ($\ell \leq 10$). Thus, we can identify 
a significant portion of the WMAP-5 observed optical depth
through known sources of ionization. 
The ``visible ionized universe" out to \zGP\ = 6--7 accounts for 
$\tau_e =$ 0.04--0.05, while high-$z$ partial ionization could contribute 
additional $\Delta \tau_e \approx$ 0.01--0.03. For this study, we 
investigate the requirements to produce an additional optical depth, 
$\Delta \tau_e = 0.03 \pm 0.02$, from star formation and early black 
hole accretion at $z >$ \zGP. 

Our calculations represent an important change in the derivation of $z_r$ 
from $\tau_e$, and suggests that the amount and efficiency of high-$z$ 
star formation need to be suppressed.  This suggestion is ironic, since WMAP-1
data initially found a high $\tau_e = 0.17 \pm 0.04$ (Spergel \etal 2003)
implying a surprisingly large redshift for early reionization, ranging from 
$11 < z_r < 30$ at 95\% C.L. (Kogut \etal 2003).  These results
precipitated many investigations of star formation at $z = 10-30$, some of
which invoked anomalous mass functions, very massive stars (VMS, with $M >
140$~\msun), and an increased ionizing efficiency from zero-metallicity
stars (VTS03; Wyithe \& Loeb 2003; Cen 2003; Ciardi, Ferrara, \& White
2003; Sokasian \etal 2003, 2004). Tumlinson, Venkatesan, \& Shull (TVS04)
disputed the hypothesis that the first stars had to be VMS.  
They showed that an IMF dominated by 10 -- 100~\msun\ stars can produce 
the same ionizing photon budget as VMS, generate CMB optical depths 
of 0.09--0.14, and still be consistent with nucleosynthetic evidence from
extremely metal-poor halo stars (Umeda \& Nomoto 2003; Tumlinson 2006; 
Venkatesan 2006).

Although the IGM recombination history, $x_e(z)$, is included in calculations 
of CMBFAST and in CMB parameter estimation, the best-fit values of $\tau_e$ 
from WMAP-5 and earlier CMB experiments have been attributed exclusively 
to the contribution from the first stars and/or black holes at $z \leq 20$. 
The contributions from post-recombination electrons ($20 < z < 1100$) have 
not always been subtracted from the data.
This post-recombination contribution was relatively small in some earlier 
models of reionization (Zaldarriaga 1997; Tegmark \& Silk 1995) that explored 
optical depths of $\tau_e = 0.5-1.0$ and suggested reionization epochs up to 
$z_r \sim 100$. However, with current data indicating late reionization, it 
becomes particularly important to consider contributions to $\tau_e$ 
at $z >$ \zGP,  prior to the first identified sources of light.
The latest WMAP-5 results (Hinshaw \etal 2008) find a lower $\tau_e$, but 
they also suggest more small-scale power available for reionizing 
sources, owing to higher normalization parameters, 
$\sigma_8 \approx 0.817 \pm 0.026$ and $\Omega_m h^2 \approx 0.1369 \pm 0.0037$.  
This change marks a significant increase ($\Delta \sigma_8 = +0.056$) over 
WMAP-3 data. Between WMAP-1 and WMAP-3, the decrease in $\sigma_8$ was offset 
by a reduction in spectrum tilt.  However, this index did not change 
significantly in the recent data, going from $n_s = 0.958\pm0.016$ (WMAP-3)
to $n_s = 0.960^{+0.014}_{-0.013}$ (WMAP-5).  
Alvarez \etal (2006) argued, from the lower values of $\tau_e$ and $\sigma_8$, 
that both WMAP-3 and WMAP-1 data require similar (high) stellar ionizing 
efficiencies.  Haiman \& Bryan (2006) use the lower $\tau_e$
to suggest that massive star formation was suppressed in minihalos.   
Our results on a lower $\Delta \tau_e$ make these requirements even 
more stringent, as we now quantify.  

In the next two sub-sections, we consider two scenarios for producing
partial IGM ionization at $z >$ \zGP.  The first considers ionizing UV
photons from the first massive stars, which will create small ``bubbles"
of fully ionized gas inside ionization fronts. 
The second scenario examines extended partial ionization produced by
X-rays from accreting black holes.  

\subsection*{3.1. Ionization by Hot Stars}

Semi-analytic and numerical models of reionization (Ricotti, Gnedin,
\& Shull 2002a,b; VTS03, Haiman \& Holder 2003) show that the efficiency
of ionizing photon injection into the IGM can be parameterized by the
production rate of ionizing (UV) photons from massive stars.   
In our approximate models, we compute the hydrogen ionization fraction 
as $x_e(z) = \epsilon_{\rm UV} f_b(z)/c_L(z)$, where
$\epsilon_{\rm UV} = N_{\gamma} f_* f_{\rm esc}$ and $f_b(z)$ is the 
fraction of baryons in collapsed halos.  To account for the nonlinearity
of recombinations, we define 
$c_L \equiv \langle n^2_{\rm HII} \rangle / \langle n_{\rm HII} \rangle^2$, 
the space-averaged clumping factor of ionized hydrogen.
We assume that $c_L$ is the same for H~II and He~III and
that $N_{\gamma} f_* f_{\rm esc}$ is constant with redshift.
As discussed by Gnedin \& Ostriker (1997), particularly in their
\S~3.1, clumping of the photoionized gas sharpens the 
reionization transition, with a recombination time much less than
the local Hubble time at $z \approx$ 6--8.  However, the integrated 
optical depth through this epoch is not greatly affected by clumping. 

The main effects of clumping arise when we translate the production rate 
of ionizing photons into the ionization history, $x_e(z)$.  When the local 
density of photoionized hydrogen is close to ionization equilibrium, 
$\langle n_{\rm HI} \, \Gamma_{\rm HI} \rangle \approx
\langle \alpha_H^{(A)}(T) \, n_e \, n_{\rm HII} \rangle$, and the  
quadratic dependence of recombination rates reduces
the local ionization fraction by a factor $c_L^{-1}$.   
When the medium is out of equilibrium, either ionizing or recombining,
this approximation, $x_e(z) = \epsilon_{\rm UV} f_b(z)/c_L(z)$, is less 
accurate.  To assess the validity of our approximation, we have run
a set of reionization models, integrating differential equations for the
propagation of ionization fronts (VTS03) to derive $x_e(z)$. When we apply 
our approximation with $c_L = 10$, it 
closely approximates the integrated hydrogen ionization history with a
clumping factor of $c_L = 40$. These two curves are virtually 
indistinguishable for nearly all the redshifts approaching reionization. 
Thus, an artificial boost by a factor of 4 in the clumping factor would 
correct for any deviations in our approximation from the true ionization 
history. Our approximation is not dramatically different in shape from 
the true ionization history, at least from our semi-analytic calculation,
and requires at most a constant multiplicative correction in the 
clumping factor.  We are currently pursuing numerical experiments
to understand the relationship between clumping and reionization,  
using adaptive mesh refinement simulations (Hallman \etal 2007) and 
exploring different weighting schemes for clumping on the sub-grids. 

The free parameters in this model are encapsulated in the 
``triple product", $f_* N_{\gamma} f_{\rm esc}$, where $f_*$ represents 
the star-formation efficiency (the fraction of a halo's baryons that go into
stars), $N_{\gamma}$ is the number of ionizing photons produced per baryon
of star formation, and $f_{\rm esc}$ is the fraction of these ionizing
photons that escape from the halo into the IGM.  We can now use our
calculations to constrain the amount of high-$z$ star
formation through the product of these three parameters.  We solve for 
the ionization history, using our previous formalism (VTS03) 
in which the baryon collapse history, $f_b(z)$, is computed through the 
Press-Schechter formalism with the cosmological parameters from WMAP-5.  
The propagation of ionization fronts is followed through the production 
rate of ionizing photons minus recombinations.

Each of these three parameters has some dependence on the halo 
mass and environment (Haiman \& Bryan 2006; Ricotti \& Shull 2000). 
Since we have already parameterized the intrahalo recombinations
through $f_{\rm esc}$, we account for the loss of ionizing photons on 
IGM scales through $c_L$ in two forms:  (1) a power-law form with slope 
$\beta = -2$ from the semi-analytic work of Haiman \& Bryan (2006); and
(2) the numerical simulations of Kohler, Gnedin \& Hamilton (2007), using their 
case C (overdensity $\delta \sim 1$ for the large-scale IGM) for $C_R$, 
the recombination clumping factor corresponding to our definition of $c_L$.
In the latter case, the clumping factor is almost constant ($c_L \approx 6$) 
until the very end of reionization. Together, these two different cases 
provide bounds on the range of possible values, although they are quite
similar over the epoch $z \approx$ 6--8 that dominates the scattering.
In our calculations, we use the Haiman \& Bryan (2006) formulation of
$C_L(z)$. 

With these assumptions, we can use equation (5) and the allowed 
additional optical depth, $\Delta \tau_e \leq 0.03$, 
to constrain the ionizing efficiency of the first stars. In Figure 1, 
we plot curves of efficiency factor, $\epsilon_{\rm UV}$, as a 
function of the resulting $\Delta \tau_e$.  The two panels illustrate
the changes in required efficiency produced by the WMAP-5 increase in
``small-scale power" compared to WMAP-3 parameters.  These differences 
arise primarily from the amplitude of fluctuations, $\sigma_8$, which 
increased from $\sigma_8 = 0.761^{+0.049}_{-0.048}$ (WMAP-3) 
to $\sigma_8 = 0.817 \pm 0.026$ (WMAP-5). In each panel, we show two curves, 
corresponding to star formation in \Lya-cooled halos, with virial 
temperature $T_{\rm vir} \approx 10^4$~K, and in H$_2$-cooled 
minihalos ($T_{\rm vir} \approx 10^3$ K).   For these calculations, we adopt
the clumping factors from Haiman \& Bryan (2006), a star-formation 
efficiency $f_\star = 0.1$, and a range of escape fractions
$f_{\rm esc} = $ 0.1--0.4,  The two curves (red and blue) illustrate 
cases designated as Population II (metal-enriched) and Population~III 
(zero-metal) stars.  Both models adopt Salpeter initial mass functions (IMF),
for which we find: 
(1) $N_{\gamma} = 60,000$ for a metal-free IMF (10--140 $M_\odot$) that 
agrees with both CMB and nucleosynthetic data (TVS04); and 
(2) $N_{\gamma} = 4000$ for a present-day IMF (1-100 $M_\odot$). 
Note that $N_{\gamma} = 34,000$ is consistent with 
$\tau_e \approx 0.084 \pm 0.016$ (WMAP-5) if photons arise from zero-metal stars 
(Tumlinson 2006).  Values of $N_{\gamma}$ were derived (TVS04) from 
the lifetime-integrated ionizing photon production from various stellar 
populations and IMFs and used as inputs in reionization models. 

Figure 1 shows how the required additional optical depth, 
$\Delta \tau_e = 0.03 \pm 0.02$, translates into the efficiency required 
of massive stars forming in high-$z$ halos.  These plots are meant to
be used as indicators of the types of star-forming halos and the
stellar populations that generate ionizing photons, per baryon that
passes through stars.  The two panels also illustrate the sensitivity of 
the efficiencies to cosmological parameters (WMAP-3 vs. WMAP-5), primarily
the amplitude of fluctuations, $\sigma_8$, which sets the small-scale
power of halos at a given epoch.   The interpretation of Figure 1 
comes from the intersection of the two rising curves (red for massive
halos, blue for mini-halos) with the two green bands, which mark the
range of expected efficiencies of Pop~II and Pop~III stars.   

One can draw several general conclusions from Figure 1. 
First, the higher amplitude of small-scale power (larger $\sigma_8$) 
in the WMAP-5 results (right panel) shifts both red and blue curves
to the right.  Lower efficiencies of ionizing-photon production
are required to produce a given additional optical depth, $\Delta \tau_e$.   
Second, considering just the WMAP-5 parameters (right panel), we see that
``massive halos", defined as those with $T_{\rm vir} \approx 10^4$~K, can 
produce optical depths ranging from $\Delta \tau_e =$ 0.005--0.018, for 
a range of Pop~III efficiencies $\epsilon_{\rm UV} =$ 600--2000.  If one 
invokes more plentiful mini-halos ($T_{\rm vir} \approx 10^3$~K), 
reionization by Pop~II stars can produce a similar 
$\Delta \tau_e =$ 0.004--0.016.  
These results suggest that partial reionization at $z >$ \zGP\
could be achieved by minihalos ($T_{\rm vir} \approx 10^3$~K), 
with a mixture of Pop~III and Pop~II stars, or a transition from one to
the other, for which the rising blue curve 
gives $\Delta \tau_e \approx$ 0.02--0.06 for
$f_{\rm esc} =$ 0.1--0.4. For these minihalos, Pop II stars produce
$\tau \la 0.02$, but $\tau \approx$ 0.02--0.06 can be achieved either
by metal-poor stellar populations in halos with unusually high
$f_*$ or $f_{\rm esc}$, or by metal-free massive stars in halos with
very low star formation rates and/or escape fraction of ionizing radiation.

Another robust conclusion from Figure 1 is that mini-haloes 
($T_{\rm vir} = 10^3$~K) with Pop~III stars can easily account for the 
{\it entire} additional optical depth ($\Delta \tau_e \approx 0.03$) needed
to explain the WMAP-5 data.  Even in cases where reionization begins with
Pop~III stars (efficiencies $\sim 10^3$) and then transitions to 
Pop~II (efficiencies $\sim 10^2$), the optical depths are significant.   
This is an important astrophysical issue, since the duration of the
epoch dominated by metal-free massive first stars is still uncertain
(TVS04).  These results mark a significant departure from similar curves 
for a WMAP-3 cosmology (left  panel), owing to the increased 
small-scale power in a  WMAP-5 cosmology relative to WMAP-3.  Although 
the scalar spectral  
index $n_s$ did not change appreciably from year 3 to year 5 of WMAP,  
both the matter density, $\Omega_m$, and the normalization, $\sigma_8$, 
rose by several percent, an appreciable effect for power available for 
low-mass halos. Thus, reionization occurs slightly earlier in a WMAP-5  
cosmology ($\Delta z_r \approx$ 1--2) in our models, lowering the required  
efficiencies by a factor of about 4.

Our predictions can be compared to those of Haiman \& Bryan (2006), who 
used WMAP-3 data to constrain the efficiency factor, $\epsilon_{\rm UV}$,
relative to a fiducial value $\epsilon_{\rm mini} = 200$ for minihalos.
They argued that the efficiency for the production of ionizing photons
must have been reduced by an order of magnitude,
in order to avoid overproducing the optical depth.  However, their 
constraints were based on the WMAP-3 value, $\tau_e \approx 0.09$,
whereas half that scattering is accounted for from the fully ionized IGM
at $z <$ \zGP.  In our formalism (Figure 1b), the efficiency constraints 
are less severe, and Pop~III minihalos can easily produce 
$\Delta \tau_e \approx 0.06$ with the expected efficiencies,
$\epsilon_{\rm UV} \approx 10^3$. 
Much larger efficiencies, close to those of metal-free stellar populations,  
and large values of $f_*$ and $f_{\rm esc}$ are required for larger halos
to produce $\Delta \tau_e \approx 0.02$.   
It makes some difference what form is assumed for the clumping factor, 
$c_L$, in constraining the ionizing efficiency of the first stars.  
The level of shifts in the (red and blue) efficiency curves is
comparable to the shifts between WMAP-3 and WMAP-5.

\subsection*{3.2. Ionization by Accreting Black Holes}

The CMB optical depth also constrains the level of X-ray preionization 
from high-redshift black holes.  Ricotti \etal (2005) were able to produce 
large optical depths, up to $\tau_e \approx 0.17$, using accreting high-$z$ 
black holes with substantial soft X-ray fluxes.  Their three simulations 
(labeled M-PIS, M-SN1, M-SN2) produced hydrogen preionization fractions 
$x_e = 0.1-0.6$ between $z = 15$ and $z = 10$, with large co-moving rates 
of star formation, $(0.001-0.1)M_{\odot}$ Mpc$^{-3}$~yr$^{-1}$, and baryon 
fractions, $\omega_{\rm BH} \approx 10^{-6}$ -- $10^{-5}$ accreted onto black 
holes.  Any significant contribution to $\tau_e$ from X-ray preionization 
requires $x_e \geq 0.1$.  
Therefore, the lower value of $\tau_e$ from WMAP-5 reduces the allowed X-ray 
preionization and black-hole accretion rates significantly  
compared to these models.  Ricotti \etal (2005) find that the IGM at 
$z >$ \zGP\ was highly ionized ($x_e \gg 0.01$).  In this limit, there were 
few X-ray secondary electrons, and most of the X-ray energy went into heating 
the ionized medium.  By contrast, the WMAP-5 optical depth suggests that 
the IGM was much less ionized at $z >$ \zGP.

Relating the effects of X-ray ionization from early black holes to a 
$\Delta \tau_e$ and a related X-ray production efficiency factor is 
less straightforward compared to the star-formation case, for the 
following reasons.  First, unlike ionization by 
UV photons, X-ray ionization is non-equilibrium in nature, and the 
timescales for X-ray photoionization at any epoch prior to $z = 6$
typically exceeds the Hubble time at those epochs (VGS01). Therefore, 
it is more difficult to establish a one-to-one correspondence between 
X-ray production at an epoch and the average efficiency of halos. 
In addition, X-ray 
ionization (whether from stars or black holes), at least initially 
when $x_e < 0.1$, will be dominated by secondary ionizations 
from X-ray-ionized helium electrons (VGS01) rather than from direct 
photoionization. This may constrain the physical conditions 
in the IGM (e.g., the level of He ionization) rather than those in the 
parent halo, when we attempt to translate a $\Delta \tau_e$ into an X-ray 
ionization efficiency. Thus, it may be difficult to make precise inferences
about the black hole density and accretion history from $\tau_e$.
 
We can define an X-ray ionization rate and efficiency parameter, 
analogous to the previous case for massive star formation.  We define 
$x_e(z) = \epsilon_X f_b(z)$, with no clumping factor, 
where $\epsilon_X = N_X f_{\rm BH} f_{\rm esc}$
and $f_b(z)$ is the fraction of baryons in collapsed halos.   
Here, $f_{\rm BH}$ is the average fraction of baryons in black holes 
in halos at $z \ga 7$ and $N_X$ is the number of X-ray ionizations of the IGM
produced by X-rays and secondary ionizations, per baryon accreted 
onto such black holes.  As we now describe, 
$f_{\rm BH}$ may approach $10^{-4}$, and standard accretion estimates
give $N_X \approx 10^6$ photons/baryon, so that $\epsilon_X \approx 100$.  

Ricotti \& Ostriker (2004) and Ricotti \etal (2005) define a parameter
$\omega_{\rm BH} = 10^{-7}$ to $10^{-6}$, equal to the fraction of baryons
that go into {\it seed} black holes. Current observations
(e.g., H\"aring \& Rix 2004) suggest that the BH-to-bulge mass fraction 
in modern galaxies is $M_{\rm BH}/M_{\rm bulge} \approx 1.4 \times 10^{-3}$.
Thus, if the bulge-to-halo mass ratio is 0.065, $f_{\rm BH} \approx 10^{-4}$,
after BHs have grown from the initial fraction through accretion, thereby
producing X-rays ionization at high-$z$.
To estimate $N_X$, we assume Eddington-limited accretion at 
10\% efficiency and adopt a mean photon energy $\sim1$ keV:
\begin{equation}
   N_X \approx \left[ \frac {0.1 m_p c^2} 
      {(1.6 \times 10^{-9}\;{\rm erg}) \, E_{\rm keV}} \right] (12) 
      \approx (10^6 \; {\rm photons/baryon}) E_{\rm keV}^{-1}  \; .  
\end{equation}
Here, we assume that each X-ray photon produces $\sim 12$ hydrogen ionizations, 
primarily through secondary ionizations from X-ray photoelectrons (Shull 
\& van Steenberg 1985; VGS01).  In the partially ionized IGM, the free
electrons come from H$^+$, He$^+$, and trace He$^{+2}$. 
We assume that the clumping factor ($c_L$)
and escape fraction ($f_{\rm esc}$) are roughly unity for X-rays, given 
their high penetrating power relative to UV photons. 
Figure 2 shows the allowed additional optical depth, analogous to the 
constraints of Figure 1, for X-ray efficiency in both \Lya-cooled 
halos and minihalos. Evidently, X-rays from black holes located in 
high-redshift minihalos  can produce $\Delta \tau_e \approx 0.02$,
with a substantial ionizing efficiency, $\epsilon_X \approx 100$, that
rivals that of Pop~II star formation. 

In summary, we have shown that the revised (WMAP-5) values of CMB optical 
depth, $\tau_e = 0.084 \pm 0.016$, lead to a more constrained picture of 
early reionization of the IGM.  Approximately half of the observed $\tau_e$ 
comes from a fully ionized IGM back to \zGP\ $\approx$ 6--7.    
The additional $\Delta \tau_e$ at $z >$ \zGP\ probably arises from the first 
massive stars and from accretion onto early black holes.  Some of the 
observed $\tau_e$ may come from scattering from residual electrons left 
from recombination; inaccuracies in computing this ionization
history add systematic uncertainty to the CMB inferred signal.  
We have assumed extra scattering, $\Delta \tau_e = 0.03 \pm 0.02$
at $z >$ \zGP\  and used this to constrain the efficiencies for
production and escape of ionizing photons, either by ultraviolet
photons from the first massive stars (Fig.\ 1) or by X-rays 
from accretion onto early black holes (Fig.\ 2).  

In both cases, the picture is of a partially ionized IGM at redshifts
$z = 6-20$.  For X-ray pre-ionization by early black holes, 
equation (6) can be used to provide an estimate of the effects of partial 
reionization. Between redshifts $z_2 \approx 20$ and $z_1 \approx 6$, 
an IGM with ionized fraction $x_{e,0} = 0.1$  (Ricotti \etal 2005)
would produce $(\Delta \tau_e) = (0.018) (x_{e,0} / 0.1)$,  
which is a significant contribution to the observed $\tau_e = 0.084 \pm 0.016$. 
Ricotti \etal (2005) suggested a large $x_e \approx 0.1-0.6$ and 
pushed their black-hole space densities and accretion rates to large values  
in order to reach the WMAP-1 estimates of $\tau_e = 0.17$.  Because such 
large values of $\tau_e$ are no longer required, the black hole densities 
and IGM ionization fractions are likely to be considerably less.  

All these constraints depend heavily on uncertain parameterizations 
of the efficiency of star formation and ionizing photon production.  
The most sensitive of these parameters is $\sigma_8$, although the
details of the clumping factor are comparably important.
With more precise measurements of CMB optical depth
from future missions, there is hope that more stringent constraints on 
high-$z$ star formation and black-hole accretion will be possible. 
These data include additional years of WMAP observations, as well as 
the {\it Planck} mission. 

\noindent
{\bf Acknowledgements} 

\noindent
We are grateful to David Spergel, Licia Verde, Rachel Bean, and Nick Gnedin 
for useful discussions regarding the interpretation of WMAP data and
numerical simulations.  We thank Douglas Scott and Wan Yan Wong for 
providing their calculations of recombination history.   
This research at the University of Colorado was supported by astrophysical 
theory grants from NASA (NNX07-AG77G) and NSF (AST07-07474).

\clearpage

\begin{figure}
  \epsscale{1.1}
  \plottwo{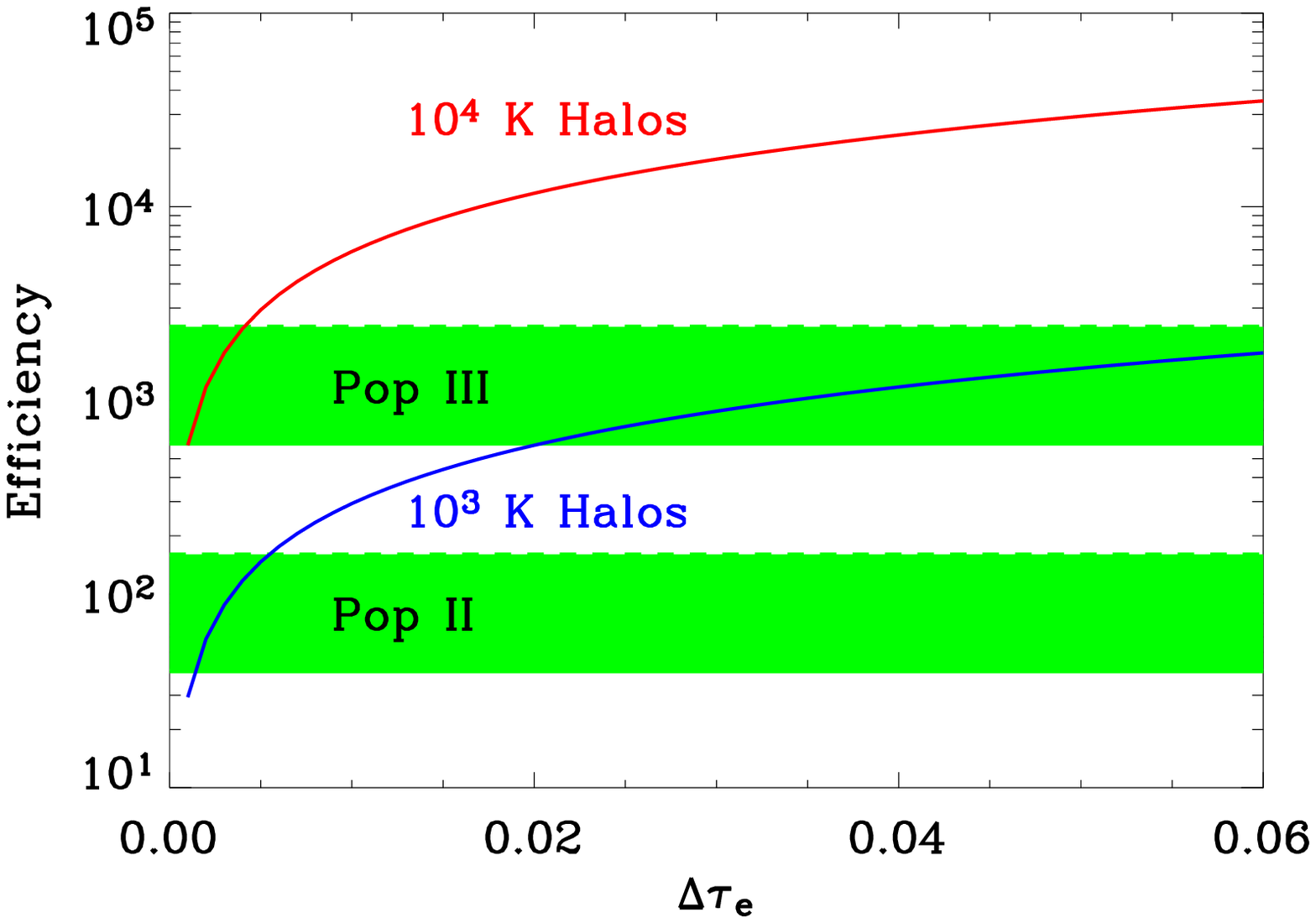}{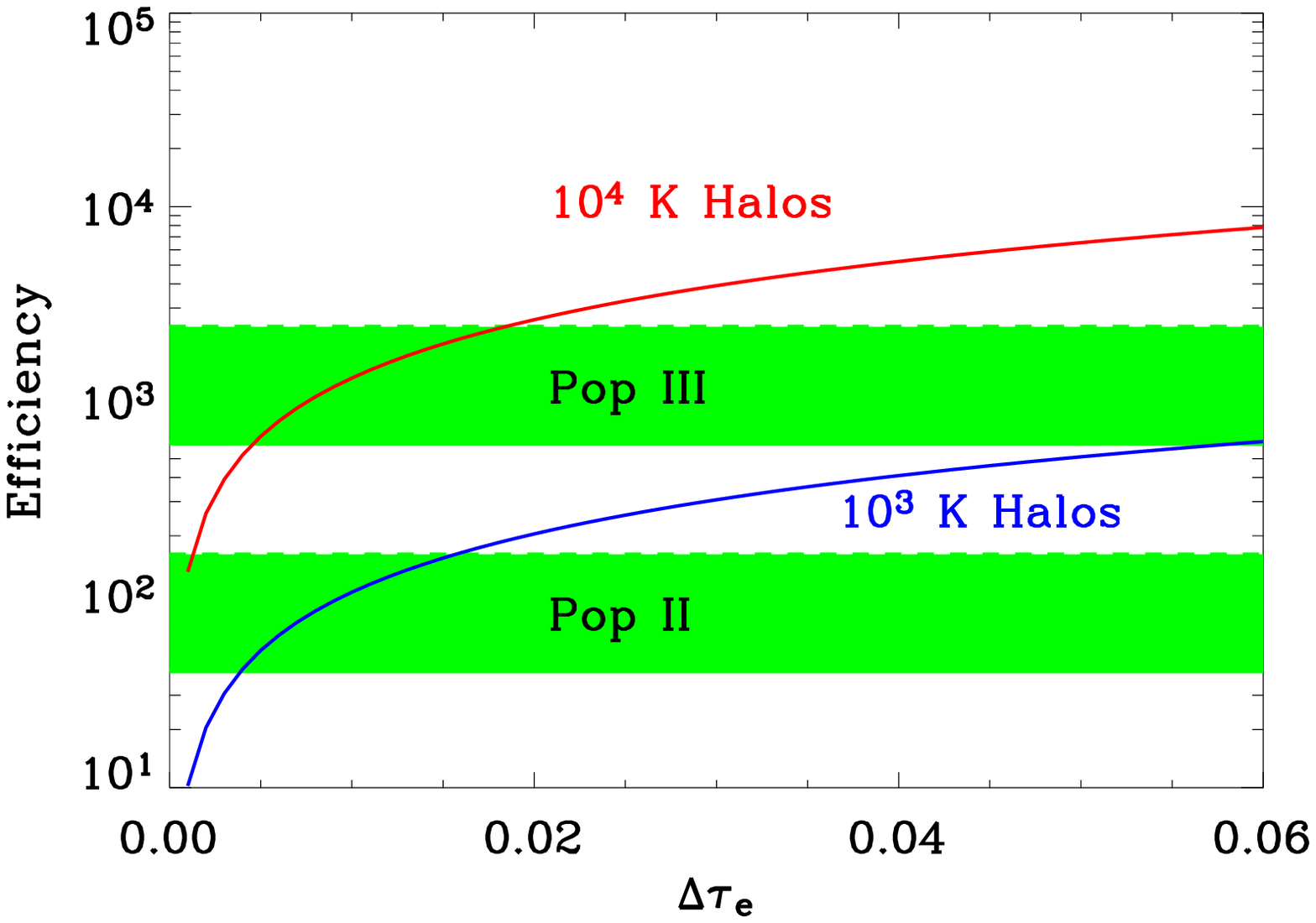} 
  \caption{Locus of efficiency factors,  
   $\epsilon_{\rm UV} = N_{\gamma} f_* f_{\rm esc}$, for production
   and escape of photoionizing radiation vs.\ additional optical depth, 
   $\Delta \tau_e$, from hot stars at $z > z_{\rm GP}$.  Models assume 
   cosmological parameters from WMAP-3 (left panel) and WMAP-5 (right panel).  
   WMAP-5 found more small-scale power (higher $\sigma_8$) for mini-halo 
   ionization, hence lower required efficiencies.  Efficiency,
   $\epsilon_{\rm UV} = f_* N_{\gamma} f_{\rm esc}$, is defined (\S~3.1) 
   for star formation in 
   \Lya-cooled massive halos (red curve, $T_{\rm vir} \geq 10^4$ K)
   and in H$_2$-cooled minihalos (blue curve, $T_{\rm vir} \geq 10^3$ K). 
   Green horizontal bands correspond to expected efficiencies for 
   a Salpeter stellar IMF: $\epsilon_{\rm UV} \approx$ 600--2000 for metal-free 
   Pop~III (10--140 $M_\odot$) and $\epsilon_{\rm UV} \approx$ 40--150 for
   metal-enriched Pop~II (1--100 $M_\odot$).  We adopt star-formation
   efficiency $f_* = 0.1$ and photon escape fraction $f_{\rm esc}$ ranging 
   from 0.1--0.4 (see \S~3.1 for details).  The intersection of the rising 
   red curve (right panel) with the upper green band shows that Pop~III halos 
   ($T_{\rm vir} \approx 10^4$~K) can produce $\Delta \tau_e \approx$ 
   0.01--0.02.  The intersection of the rising blue curve with the two 
   bands shows that mini-halos ($T_{\rm vir} \approx 10^3$~K) can produce 
   $\Delta \tau_e =$ 0.01--0.02 (Pop~II) and $\Delta \tau_e$ up to 0.06 
   (Pop~III/Pop~II). }    
\end{figure}

\clearpage

\begin{figure}
  \epsscale{1.1}
  \plottwo{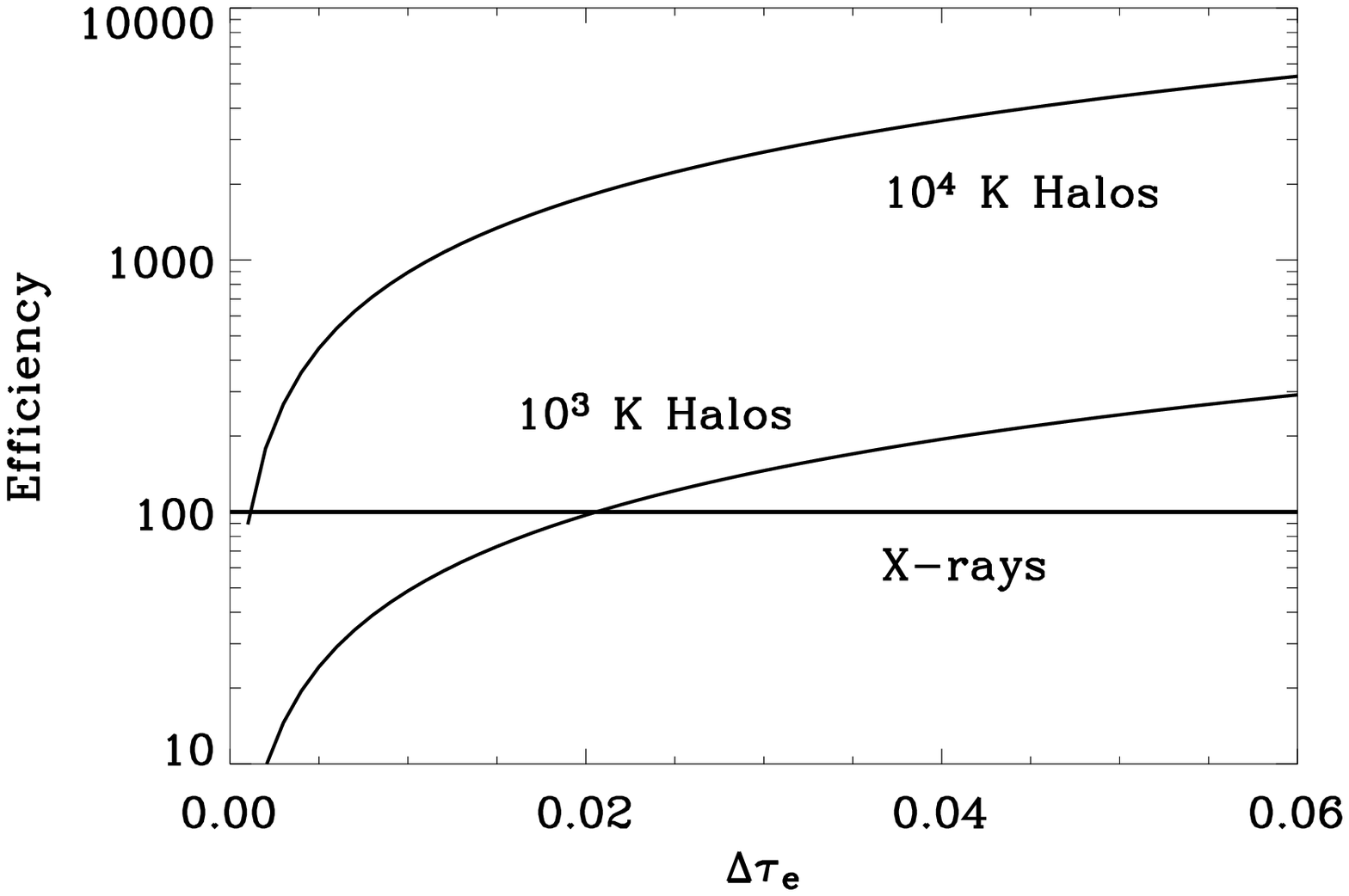}{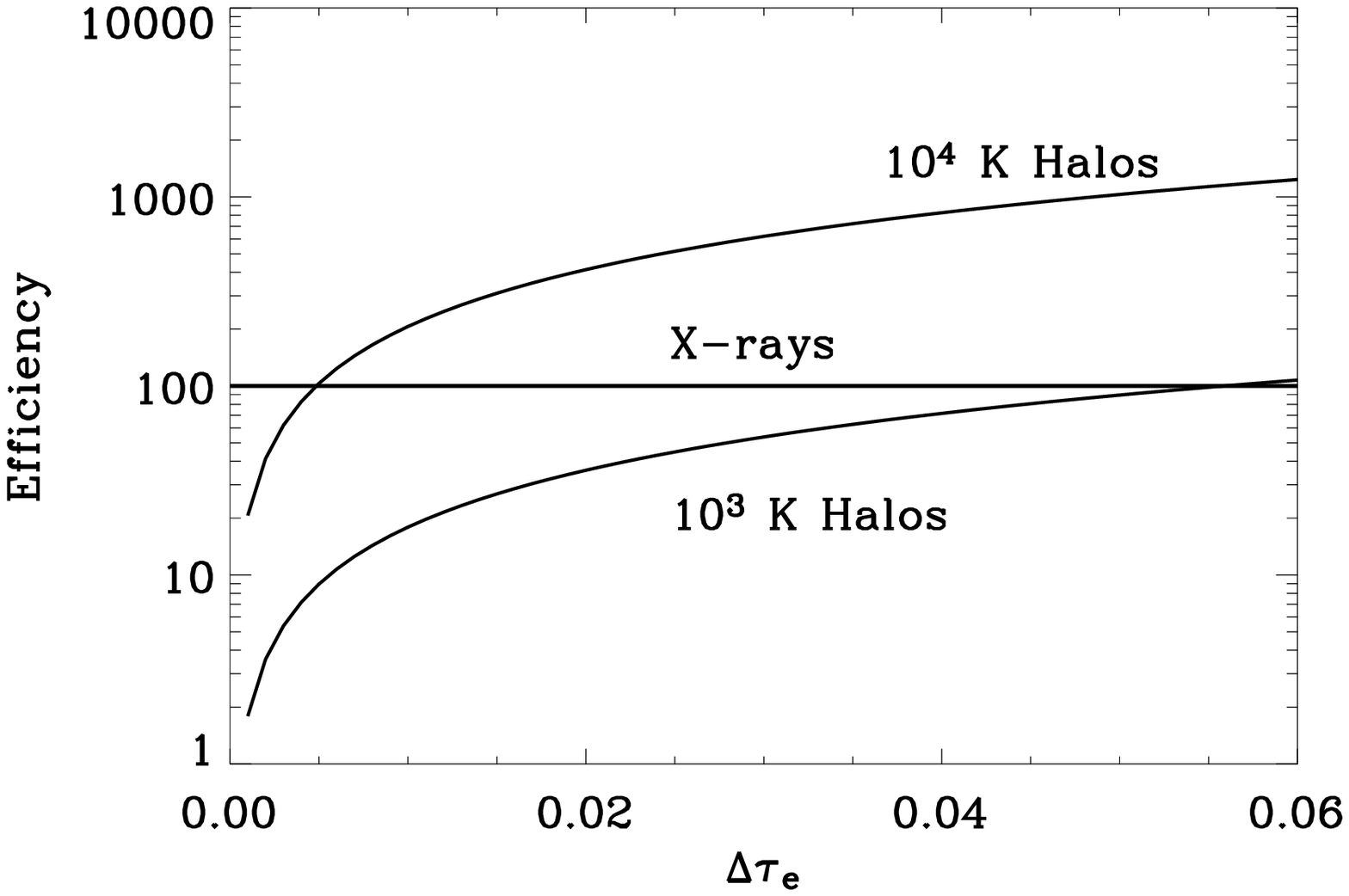}
  \caption{X-ray production efficiency, $\epsilon_X = N_X f_{\rm BH} f_{\rm esc}$,  
    required to produce additional optical depth, $\Delta \tau_e$, at 
    $z > z_{\rm GP}$.  We assume that $f_{\rm esc} = 1$ for X-rays.  
    This efficiency is the product of the baryon fraction in black holes
    times the total ionizations produced per accreted baryon.  
    Curves are for WMAP-3 parameters (left) and WMAP-5 (right).
    The required efficiencies are lower for WMAP-5 parameters, with their 
    additional small-scale power.   Two curves are 
    shown, for halos with virial temperatures of $10^3$~K and $10^4$~K. 
    Horizontal curves show a fiducial value, $\epsilon_X \approx 100$, 
    equivalent to $f_{\rm BH} \approx 10^{-4}$, $N_X \approx 10^6$ 
    (see \S~3.2), and $f_{\rm esc} = 1$. } 
\end{figure}

\end{document}